\documentclass[12pt]{article}
\begin{document}

\newcommand{\beq}{\begin{equation}}
\newcommand{\eeq}[1]{\label{#1}\end{equation}}
\newcommand{\bea}{\begin{eqnarray}}
\newcommand{\eea}[1]{\label{#1}\end{eqnarray}}
\newcommand{\Tr}{\mbox{Tr}\,}
\newcommand{\bm}[1]{\mbox{\boldmath{$#1$}}}

\begin{titlepage}

\begin{center}
{\Large \bf Kerr quasinormal modes and Hod's time-temperature bound. }

\vspace{20pt}

{\large A. Gruzinov}

\vspace{12pt}

{ CCPP\\ 
Department of Physics\\ New York University\\
4 Washington Pl.\\ New York, NY 10003}
\end{center}

\vspace{20pt}

\begin{abstract}
We give an explicit expression for the frequencies of slowly damped quasinormal modes of near-extreme Kerr black holes. It follows from this expression that the near-extreme Kerr holes obey the Hod's bound: in the limit of maximal rotation, $\lim \sup \omega _{IS}/T\leq \pi / \hbar$, where $\omega _{IS}$ is the decay rate of the slowest decaying quasinormal mode, $T$ is the black hole temperature. On the other hand, the bound is not saturated in the sense that $\lim \inf \omega _{IS}/T< \pi /\hbar$ is a strict inequality. {\it It remains unclear} whether the bound is saturated in the sense that $\lim \sup \omega _{IS}/T= \pi /\hbar$.

\end{abstract}

\end{titlepage}

\newpage

\section{Introduction}

Hod \cite{hod} has recently proposed a remarkable universal bound on the relaxation time of an arbitrary system in thermal equilibrium: 
\beq
{\omega _I \over T}\leq {\pi \over \hbar },
\eeq{bound}
where $\omega _I$ is the slowest relaxation rate of the system, $T$ is the temperature. He further proposed that this bound is saturated by the near extreme black holes (black holes with $T\rightarrow 0$), if one identifies the damping rates with the (minus) imaginary part of the quasinormal mode frequencies.

The saturation property might be of interest, because the Hod's bound is obtained from the Bremermann-Bekenstein bound on the rate of information transfer, and Bekenstein's \cite{bek} derivation does not seem to give the exact dimensionless coefficient. In fact, Bekenstein \cite{bek} notes that it might be possible to obtain a tighter bound. 

To support the saturation property, Hod numerically calculated the imaginary part of the quasinormal mode frequency for the fundamental ($n=1$) scalar ($s=0$) and gravitational ($s=2$) perturbations of a Kerr hole with $l=m=2$. He noted that the decay rates do obey the bound and seem to saturate it in the extreme limit. 

Here we give an explicit expression for the frequencies of slowly damped quasinormal modes of near-extreme Kerr black holes. As follows from our expression:

\begin{enumerate}

\item Individual slowly damped fundamental modes (in particular the modes $(s=0,n=1,l=2,m=2)$ and $(s=2,n=1,l=2,m=2)$ considered by Hod) do not obey the bound (\ref{bound}).

\item The slowest decaying mode, with the decay rate $\omega _{IS}$ obeys the bound (\ref{bound}) in the sense that in the limit of extreme rotation $\lim \sup \omega _{IS}/T\leq \pi / \hbar$.

\item The slowest decaying mode does not saturate the bound in the sense that $\lim \inf \omega _{IS}/T< \pi /\hbar$ is a strict inequality.

\end{enumerate}

It might still be true that the bound is saturated in the sense that  $\lim \sup \omega _{IS}/T= \pi /\hbar$, but this remains to be seen.

\section{Slowly-damped quasinormal modes of the near-extreme Kerr hole}

\subsection{Frequencies of slowly damped modes}

We first give the expression for the frequency, then confirm it by comparison with exact (numerically) results, and finally we briefly explain its origin. We follow the definitions and notations of Teukolsky and Press \cite{teuk}.

With $G=c=1$ we take the black hole of mass $M=1$. The hole is near-extreme, meaning that the rotation parameter $a$ is close to the maximal value 1:  $\sigma \equiv (r_+-r_-)/r_+\ll 1$, where $r_{\pm }=1\pm \sqrt{1-a^2}$. 

For some values of $s$, $l$, $m$, the eigenvalue $\delta \equiv \delta _{slm}$ of the angular equation 
\beq
{1\over \sin \theta}{d\over d\theta}\left( \sin \theta {dS\over d\theta }\right)+\left( ({m\over 2}\cos \theta-s)^2-{(m+s\cos \theta)^2\over \sin ^2 \theta }+{7\over 4}m^2-2s^2-{1\over 4}-\delta ^2\right)S=0
\eeq{ang}
is real. The fundamental modes with these $s$, $l$, $m$  are slowly damped. Their frequencies are given by
\beq
\omega _{slm}(\sigma )={m\over 2}-{\sigma \over 4}\left( \delta +{i\over 2}-iz\exp (-2i\delta \ln \sigma ) \right) + o(\sigma),
\eeq{freq}
where $z\equiv z_{slm}$ is given by 
\beq
z=-{\exp (-\pi \delta -2i\delta \ln m)\Gamma(1+2i\delta)\Gamma ({1\over 2}+s-im-i\delta)\Gamma ({1\over 2}-s-im-i\delta)\over \Gamma(-2i\delta)\Gamma(1-2i\delta)\Gamma ({1\over 2}+s-im+i\delta)\Gamma ({1\over 2}-s-im+i\delta)}+o(z),
\eeq{zfreq}
where $\Gamma$ is the $\Gamma$-function.

\subsection{Numerical check}

As seen from (\ref{zfreq}), our expression (\ref{freq}) should be accurate for those $s$, $l$, $m$ which give  small $|z_{slm}|$. In practice, this means that (\ref{freq}) works for all modes with real $\delta _{slm}$. 

For example, solving the angular equation for scalar ($s=0$) perturbations, one finds that the smallest values of $l$ and $m$ with positive $\delta _{0lm}$ are $l=m=2$ with $\delta _{022}=0.94596$. Then equation (\ref{zfreq}) gives $|z_{022}|=0.001441\ll 1$, and we expect that (\ref{freq}) should be accurate. 

Take some rotation parameter for which the frequency (\ref{freq}) violates the Hod's bound. For example, for $\sigma=0.002$ equation (\ref{freq}) gives the frequency 
\beq
\omega _{022}(0.002)=0.999527263-0.000250678i,
\eeq{fnum}
while the black hole temperature (times $\pi /\hbar$)
\beq
{\pi T\over \hbar}\approx {\sigma \over 8}-{\sigma ^2 \over 16}
\eeq{temp}
is equal to 0.00024975 in this case.

We now compare our approximate value (\ref{fnum}) to the ``true value'' which we calculate using Leaver's \cite{leav} continued fraction dispersion law computed by the brute-force double precision numerics. This gives 
\beq
\omega _{``exact''~022}(0.002)=0.999527301-0.000250661i
\eeq{ftru}
in agreement with (\ref{fnum}). We, of course, checked many more cases. 

\subsection{Derivation}

Equation (\ref{zfreq}) immediately follows from the Detweiler's formula (equation (9) of \cite{detw}), once the smallness of the parameter $z$ (defined by (\ref{freq})) is assumed (the $\Gamma$-function in the denominator of the r.h.s. of (9) of \cite{detw} approaches the pole).

\section{Hod's bound}
It remains to prove the statements 1-3 of \S1. 

{\it 1 and 3 -- bound not saturated}: As seen from (\ref{freq}), (\ref{zfreq}), when $\sigma$ approaches 0 from above, the normalized damping rate ${\hbar \omega _I\over \pi T}$ makes infinitely many oscillations with constant amplitude around 1.

{\it 2 -- bound not violated}: For large values of $m$ and $\delta$ (say for $\delta_{0mm}$), equation (\ref{zfreq}) gives arbitrarily small $z$, and then (\ref{freq}) gives ${\hbar \omega _I\over \pi T}$ arbitrarily close to 1.

\subsection*{Acknowledgments}
I thank Shahar Hod and Mat Kleban for useful discussions. This work was supported by the David and Lucile Packard foundation.

%%%%%%%%%%%%%%%%%%%%%%%%%%%%%%%%%%
%%%%%%%%%%%%%%%%%%%%%%%%%%%%%%%%%%%
%%%%%%%%%%%%%%%%%%%%%%%%%%%%%%%%%%

\end{document}